# Medical Photoacoustic Beamforming Using Minimum Variance-Based Delay Multiply and Sum

Moein Mozaffarzadeh[a], Ali Mahloojifar[a], and Mahdi Orooji[a]

[a]Department of Biomedical Engineering, Tarbiat Modares University, Tehran, Iran

## ABSTRACT

Delay-and-Sum (DAS) beamformer is the most common beamforming algorithm in Photoacoustic imaging (PAI) due to its simple implementation and real time imaging. However, it provides poor resolution and high levels of sidelobe. A new algorithm named Delay-Multiply-and-Sum (DMAS) was introduced. Using DMAS leads to lower levels of sidelobe compared to DAS, but resolution is not satisfying yet. In this paper, a novel beamformer is introduced based on the combination of Minimum Variance (MV) adaptive beamforming and DMAS, so-called Minimum Variance-Based DMAS (MVB-DMAS). It is shown that expanding the DMAS equation leads to some terms which contain a DAS equation. It is proposed to use MV adaptive beamformer instead of existing DAS inside the DMAS algebra expansion. MVB-DMAS is evaluated numerically compared to DAS, DMAS and MV and Signal-to-noise ratio ($SNR$) metric is presented. It is shown that MVB-DMAS leads to higher image quality and $SNR$ for about 13 $dB$, 3 $dB$ and 2 $dB$ in comparison with DAS, DMAS and MV, respectively.

## 1. INTRODUCTION

Photoacoustic imaging (PAI) is an emerging medical imaging modality that uses short electromagnetic pulse to generate Ultrasound (US) waves based on thermoelastic effect.[1] The main motivation of using PAI is having merits of the US spatial resolution and optical imaging contrast in one modality.[2] PAI is a multiscale imaging modality that has been used in different fields of study such as tumor detection,[3,4] ocular imaging,[5] monitoring oxygenation in blood vessels[6] and functional imaging.[7,8] There are two techniques of PAI: Photoacoustic Tomography (PAT) and Photoacoustic Microscopy (PAM).[9,10] In 2002, for the first time, PAT was successfully used as *in vivo* functional and structural brain imaging modality in small animals.[11] In PAT, an array of elements may be formed as linear, arc or circular shape and mathematical reconstruction algorithms are used to obtain optical absorption distribution map of a tissue.[12–14]

Since there is a high similarity between US and PA detected signals, many of beamforming algorithms used in US imaging can be used in PAI. Moreover, integrating these two imaging modalities has been a challenge over the past few years.[15,16] Although DAS is the most common beamforming method in linear array imaging, it is a nonadaptive and blind beamformer. Consequently, DAS causes a wide mainlobe width and high levels of sidelobe. Minimum Variance (MV) can be considered as one of the commonly used adaptive methods used in medical imaging.[17,18] Over time, vast variety of modifications have been investigated on MV such as complexity reduction,[19,20] shadowing suppression,[21] make use of eigenstructure to enhance MV performance.[22,23] Matrone et al. proposed, in,[24] a new beamforming algorithm namely Delay-Multiply-and-Sum (DMAS) as a beamforming technique, used in medical US imaging. In our last work, we have developed a novel algorithm outperforming DMAS.[25]

In this paper, a new algorithm namely Minimum Variance-Based DMAS (MVB-DMAS) is proposed. It is shown that there is a DAS algebra inside the expansion of DMAS algorithm, and it is proposed to use MV instead of each term of existing DMAS. This algorithm is called Minimum Variance-Based Delay Multiply and Sum (MVB-DMAS). It is shown that using MVB-DMAS leads to higher image quality in comparison with DAS, DMAS and MV.

---

Further author information: (Send correspondence to A. Mahloojifar)
Ali Mahloojifar: E-mail: mahlooji@modares.ac.ir, +98 21 82883304

## 2. METHODS

In this section, the basic idea of the proposed algorithm and the necessary modification are explained.

### 2.1 Beamforming Algorithm

When PA signals are detected by a linear array of US transducer, US beamforming algorithms such as DAS can be used to reconstruct the image from detected PA signals, using the following equation:

$$y_{DAS}(k) = \sum_{i=1}^{M} x_i(k - \Delta_i), \tag{1}$$

where $y_{DAS}(k)$ is the output of the beamformer, $k$ is the time index, $M$ is the number of elements of array, and $x_i(k)$ and $\Delta_i$ are detected signals and corresponding time delay for detector $i$, respectively. DAS is a simple algorithm and can be used for realtime US and PA imaging. However, in linear array transducer only a few number of detection angles are available. In other words, low quality image is formed due to limited available angles in linear array transducers. To address this problem, DMAS was introduced in.[24] The same as DAS, DMAS calculates corresponding sample for each elements of the array based on delays, but before summation, samples are combinatorially coupled and multiplied. DMAS formula is given by:

$$y_{DMAS}(k) = \sum_{i=1}^{M-1} \sum_{j=i+1}^{M} x_i(k - \Delta_i) x_j(k - \Delta_j). \tag{2}$$

To overcome the dimensionally squared problem of (2) following equations are suggested:[26]

$$\hat{x}_{ij}(k) = \text{sign}[x_i(k - \Delta_i) x_j(k - \Delta_j)] \sqrt{|x_i(k - \Delta_i) x_j(k - \Delta_j)|}, \quad \text{for} \quad 1 \leqslant i \leqslant j \leqslant M. \tag{3}$$

$$y_{DMAS}(k) = \sum_{i=1}^{M-1} \sum_{j=i+1}^{M} \hat{x}_{ij}(k). \tag{4}$$

Performing sign, absolute and square root after coupling procedure in (3) and (4), which requires $(M^2 - M)/2$ computations for each pixel, result in slow imaging. To put it more simply, sometimes these library functions require many clock cycles which leads to improper timing performance of DMAS algorithm. Applying following procedure to the received PA signals reduces computational number of the sign, absolute and square root operations to $M$ for each pixel[26]

$$\bar{x}_i(k) = \text{sign}[x_i(k)] \sqrt{x_i(k)} \quad \text{for} \quad 1 \leqslant i \leqslant M, \tag{5}$$

$$\hat{x}_{ij}(k) = \bar{x}_i(k) \bar{x}_j(k) \quad \text{for} \quad 1 \leqslant i \leqslant j \leqslant M. \tag{6}$$

The procedure of DMAS algorithm can be considered as a correlation process which uses the auto-correlation of aperture. In other words, the output of this beamformer is the spatial coherence of detected PA signals, and it is a non-linear beamforming algorithm. Consider the expansion of DMAS algorithm which can be written as follows:

$$y_{DMAS}(k) = \sum_{i=1}^{M-1} \sum_{j=i+1}^{M} x_{id}(k) x_{jd}(k) = x_{1d}(k) \underbrace{\left[ x_{2d}(k) + x_{3d}(k) + x_{4d}(k) + ... + x_{Md}(k) \right]}_{\text{first term}}$$

$$+ x_{2d}(k) \underbrace{\left[ x_{3d}(k) + x_{4d}(k) + ... + x_{Md}(k) \right]}_{\text{second term}} + ... + x_{(M-2)d}(k) \underbrace{\left[ x_{(M-1)d}(k) + x_{Md}(k) \right]}_{\text{(M-2)th term}} + \underbrace{\left[ x_{(M-1)d}(k) . x_{Md}(k) \right]}_{\text{(M-1)th term}}.$$

$$\tag{7}$$

where $x_{id}(k)$ and $x_{jd}(k)$ are delayed detected signals for element $i$ and $j$, respectively. As can be seen, There is a DAS in every terms of expansion, and it can be used to modify the DMAS beamformer. In (7), in every terms, there exists a summation procedure which is a type of DAS algorithm. It is proposed to use MV adaptive beamformer for each term instead of DAS. In other words, since DAS is a non-adaptive beamformer and considers all calculated samples for each element of the array the same as each other, consequently, the acquired image by each term is a low quality image with high levels of sidelobe and broad mainlobe. In order to use MV instead of every DAS in the expansion, we need to carry out some modifications and prepare the expansion in (7) for the proposed method. Following section contains the essential modifications.

## 2.2 Modified DMAS

It should be noticed that the quality of covariance matrix estimation in MV is highly depended on the selected length of subarray. The upper band is limited to $M/2$ and the lower band to 1. In (7), each term can be considered as a representation of a DAS algorithm with different number of elements of array. Limited number of entries in each term causes problem for MV algorithm due to the limited length of the subarray. This problem can be addressed by adding the unavailable elements in each term in order to acquire large enough number of available elements and consequently high quality covariance matrix estimation. The extra terms, needed to address the problem, are given by:

$$y_{total-extra}(k) = \sum_{i=M-1}^{2} \sum_{j=i-1}^{1} x_{id}(k) x_{jd}(k) + y_{extra} = x_{(M-2)d}(k)\Big[x_{(M-3)d}(k) + x... + x_{2d}(k)\Big] + x_{1d}(k)\Big] \\ + ... + x_{3d}(k).\Big[x_{2d}(k) + x_{1d}(k)\Big] + x_{2d}(k) x_{1d}(k) + y_{extra}(k), \tag{8}$$

where

$$y_{extra}(k) = x_{Md}(k)\Big[x_{(M-1)d}(k) + x_{(M-2)d}(k) + ... + x_{3d}(k) + x_{2d}(k) + x_{1d}(k)\Big]. \tag{9}$$

MV can be implemented using all entities in each term. The expansion of MVB-DMAS can be written as follows:

$$y_{MVB-DMAS}(k) = \sum_{i=1}^{M} x_{id}(k)\big(\boldsymbol{W}_{i,M-1}^{H}(k)\boldsymbol{X}_{id,M-1}(k)\big) = \sum_{i=1}^{M} x_{id}(k)\bigg(\sum_{j=1,j\neq i}^{M} w_j(k) x_{jd}(k)\bigg) = \\ \sum_{i=1}^{M} x_{id}(k)\bigg(\boldsymbol{W}^H(k)\boldsymbol{X}_d(k) - w_i(k) x_{id}(k)\bigg) = \sum_{i=1}^{M} x_{id}(k)\bigg(\sum_{j=1}^{M} w_j(k) x_{jd}(k) - w_i(k) x_{id}(k)\bigg) = \\ \sum_{i=1}^{M} x_{id}(k) \underbrace{\bigg(\sum_{j=1}^{M} w_j(k) x_{jd}(k)\bigg)}_{MV} - \sum_{i=1}^{M} x_{id}(k)\bigg(w_i(k) x_{id}(k)\bigg), \tag{10}$$

where $\boldsymbol{X}_d(k)$ is time-delayed array detected signals $\boldsymbol{X}_d(k) = [x_1(k), x_2(k), ..., x_M(k)]^T$, $\boldsymbol{W}(k) = [w_1(k), w_2(k), ..., w_M(k)]^T$ is the beamformer weights, and $(.)^T$ and $(.)^H$ represent the transpose and conjugate transpose, respectively. $\boldsymbol{W}_{i,M-1}$ and $\boldsymbol{X}_{id,M-1}$ are almost the same as $\boldsymbol{W}(k)$ and $\boldsymbol{X}_d(k)$, respectively, but the $i_{th}$ element of the array is ignored in calculation and as a result, the length of these vectors becomes $M-1$ instead of $M$.

## 3. RESULTS

K-wave Matlab toolbox was used to simulate the numerical study.[27] Thirty 0.1 $mm$ radius spherical absorbers as initial pressure were positioned along vertical axis every 5 $mm$ beginning 25 $mm$ from transducer surface while there are three absorbers at each depths of imaging. Imaging region was 40 $mm$ in lateral axis and 75 $mm$ in vertical axis. Linear array having $M$=128 elements operating at 5 $MHz$ central frequency and 77% fractional bandwidth was used to detect the PA signals generated from defined initial pressures. Speed of sound was assumed to be 1540 $m/s$ during simulations. Sampling frequency was 50 $MHz$, subarray length $L=M/2$, $K$=5 and $\Delta = 1/100L$ for all simulations.

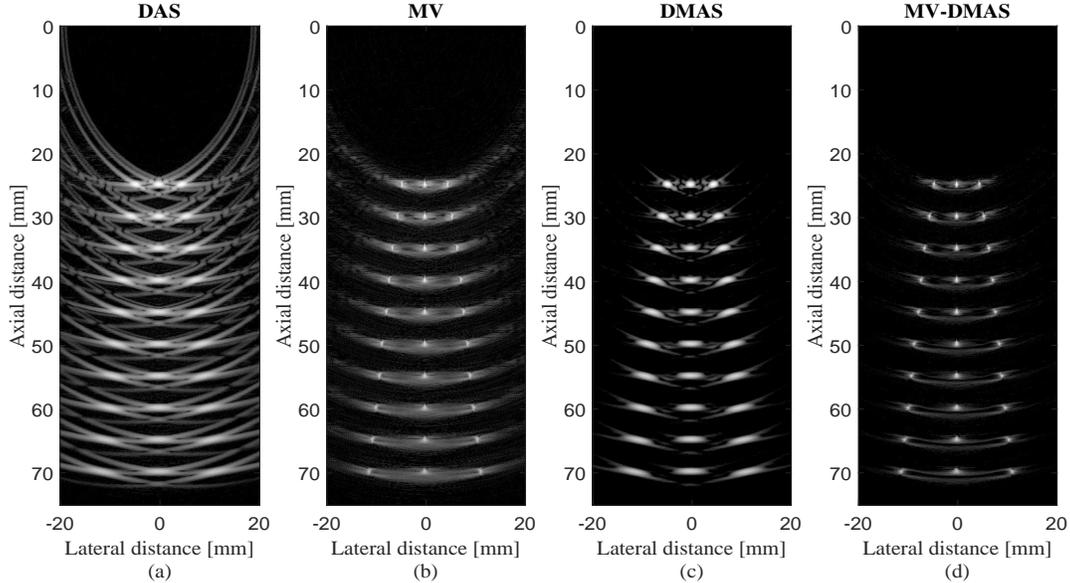

Figure 1: Simulated point targets using linear array. (a) DAS, (b) MV, (c) DMAS, (d) MVB-DMAS. All images are shown with a dynamic range of 60 $dB$. 50 $dB$ noise was added to detected signals.

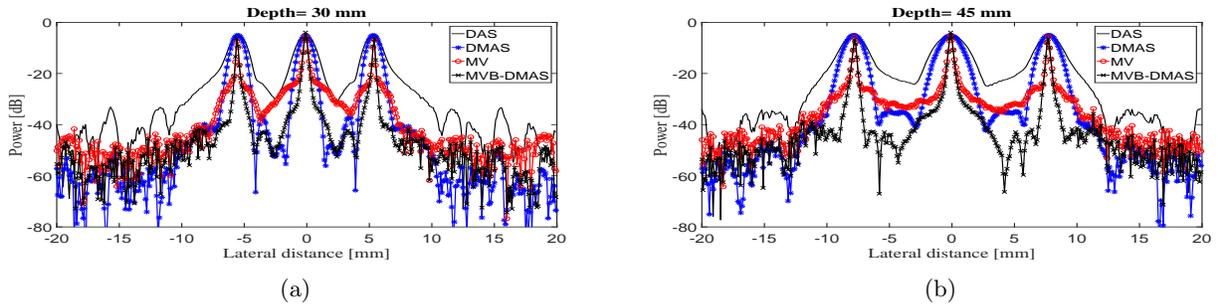

Figure 2: Lateral variation of DAS, MV, DMAS and MVB-DMAS in depth of (a) 30 $mm$ and (b) 45 $mm$ while sound velocity is 5% overestimated.

The reconstructed images using mentioned beamformers are shown in Figure 1. As can be seen, the reconstructed image using DAS contains high levels of sidelobe and artifacts, and the quality of reconstructed image is low. MV improves the quality of the reconstructed image in comparison with DAS. However, the levels of sidelobe are annoying and degrading the quality of image. The formed image using DMAS is shown in Figure 1(c), and it can be seen that DMAS results in lower levels of sidelobe compared to DAS and MV. However, the resolution of point targets are still low. Using MVB-DMAS as a beamforming algorithm leads to the low levels of sidelobe of DMAS and the high resolution of MV beamformer. To compare the reconstructed images in detail, lateral variation of the beamformers are shown in Figure 2. As can be seen, MVB-DMAS leads to lower levels of sidelobe and higher resolution compared to DAS, DMAS and MV. The width of mainlobe for MVB-DMAS is similar to MV and lower than DMAS, but the levels of sidelobe is lower than MV. Moreover, the lateral valley of MVB-DMAS is lower compared to MV. The lateral variations in the two depths of imaging shows the superiority of MVB-DMAS in comparison with other beamformers.

To compare the proposed method quantitatively, the proposed method is evaluated using Signal-to-Noise Ratio ($SNR$), and the calculated $SNR$s are presented in Table 1 for three depths of imaging. As can be seen, MVB-DMAS results in higher $SNR$ in all the three depths of imaging. Consider, in particular, the depth of 50 $mm$ where MVB-DMAS outperforms DAS, DMAS and MV for about 13 $dB$, 3 $dB$ and 2 $dB$, respectively.

Table 1: SNR (*dB*) Values in Different Depths.

| Depth(*mm*) | DAS | DMAS | MV | MVB-DMAS |
|---|---|---|---|---|
| 30 | 33.7381 | 50.3507 | 53.5639 | 56.3411 |
| 50 | 30.3199 | 42.0305 | 46.8804 | 48.1118 |
| 70 | 28.2905 | 38.6430 | 39.2435 | 41.8570 |

## 4. CONCLUSION

DAS beamformer is the most commonly used beamforming algorithm in US and PA imaging. However, it leads to low quality images along with a low resolution. DMAS was introduced to address the problems in DAS. However, it leads to low resolution images. Moreover, the resolution improvement is significantly high using MV. In this paper, we proposed to combine DMAS and MV to have a beamformer providing both high resolution and low levels of sidelobe. MVB-DMAS is based on the expansion of DMAS algorithm, and is was proved that it leads to higher $SNR$, in compare to other beamformers. Moreover, it results in higher resolution and lower levels of sidelobe in comparison with DMAS and MV, respectively.